\documentclass[letterpaper]{article} 
\usepackage[]{aaai25}  
\usepackage{times}  
\usepackage{helvet}  
\usepackage{courier}  
\usepackage[hyphens]{url}  
\usepackage{graphicx} 
\usepackage{natbib}  
\usepackage{caption} 
\usepackage{algorithm}
\usepackage{algorithmic}

\usepackage{newfloat}
\usepackage{listings}
\DeclareCaptionStyle{ruled}{labelfont=normalfont,labelsep=colon,strut=off} 
\floatstyle{ruled}
\newfloat{listing}{tb}{lst}{}
\floatname{listing}{Listing}
\usepackage[utf8]{inputenc} 
\usepackage{amsmath,cite,url}
\usepackage{graphicx}
\usepackage{color}

\usepackage{caption}
\usepackage{subcaption}

\usepackage{array}
\usepackage{tabularx}

\usepackage{booktabs}
\usepackage{tikz}
\usetikzlibrary{shapes.misc, positioning}

\colorlet{myred}{red!80!black}
\colorlet{myblue}{blue!80!black}
\colorlet{mygreen}{green!60!black}
\colorlet{myyellow}{yellow!60!black}
\colorlet{myorange}{orange!90!black}
\colorlet{mypurple}{purple!60!black}
\colorlet{mypink}{magenta!90!black}
\colorlet{mydarkred}{myred!40!black}
\colorlet{mydarkblue}{myblue!40!black}
\colorlet{mydarkgreen}{mygreen!40!black}
\colorlet{mygray}{gray!40!black}
\colorlet{myblack}{black}

\newcommand{\TT}[1]{$\texttt{#1}$}

\newcommand{\systemName}{MIDI-GPT}
\newcommand{\systemNameLong}{MIDI-GPT}
\newcommand{\stylerank}{StyleRank}
\newcommand{\PP}[1]{P_{#1}(\sigma)}

\title{\systemNameLong: A Controllable Generative Model for Computer-Assisted Multitrack Music Composition}
\author {
    Philippe Pasquier\textsuperscript{\rm 1},
    Jeff Ens\textsuperscript{\rm 1},
    Nathan Fradet\textsuperscript{\rm 1},
    Paul Triana\textsuperscript{\rm 1},
    Davide Rizzotti\textsuperscript{\rm 1},\\
    Jean-Baptiste Rolland\textsuperscript{\rm 2},
    Maryam Safi\textsuperscript{\rm 2}
}
\affiliations {
    \textsuperscript{\rm 1}Metacreation Lab - Simon Fraser University, Vancouver, Canada\\
    \textsuperscript{\rm 2}Steinberg Media Technologies GmbH, Hamburg, Germany\\
    pasquier@sfu.ca
}

\nocopyright

\begin{document}

\maketitle

\begin{abstract}
We present and release MIDI-GPT, a generative system based on the Transformer architecture that is designed for computer-assisted music composition workflows. MIDI-GPT supports the infilling of musical material at the track and bar level, and can condition generation on attributes including: instrument type, musical style, note density, polyphony level, and note duration. In order to integrate these features, we employ an alternative representation for musical material, creating a time-ordered sequence of musical events for each track and concatenating several tracks into a single sequence, rather than using a single time-ordered sequence where the musical events corresponding to different tracks are interleaved. We also propose a variation of our representation allowing for expressiveness. We present experimental results that demonstrate that MIDI-GPT is able to consistently avoid duplicating the musical material it was trained on, generate music that is stylistically similar to the training dataset, and that attribute controls allow enforcing various constraints on the generated material. We also outline several real-world applications of MIDI-GPT, including collaborations with industry partners that explore the integration and evaluation of MIDI-GPT into commercial products, as well as several artistic works produced using it.
\end{abstract}

%

\section{Introduction}\label{sec:introduction}

Recent research on generative music systems \cite{huang2018music, huang2020pop, briot2017deep, bpe-for-symbolique-music2023, impact-time-note-dur-sm2023} has mainly focused on modeling musical material as an end-goal, rather than on their affordance in practical scenarios \cite{sturm2019machine}. Although these works pave the way for efficient generative methods for music, their usability in real-world co-creative conditions remains limited. As a result, while we have seen a wide adoption of generative models for language and vision tasks, this has not occurred to the same extent for symbolic music composition. In contrast, artists recently expressed their concerns about the misuse of artificial intelligence in the music field \cite{artists-rights-alliance}.
If we want musicians to adopt generative systems, we must work on making models controllable, able to generate content that the user will appropriate as theirs, and integrated into their existing workflows.

Motivated by these considerations, we introduce \systemName, a style-agnostic generative model that builds on an alternative representation for multi-track musical material \cite{pasquier2018mmm}, resulting in an expressive and steerable generative system. We outline the ongoing real-world usage of \systemName\, and provide quantitative evidence to support our claim that \systemName\, rarely duplicates the training data as the length of the generated material increases; generates musical material that retains the stylistic characteristics of the training data; and that attribute control methods are an effective way to steer generation.

\section{Background}

Considering our interest in developing a system that is well suited to practical and interactive computer-assisted composition applications, we must identify the factors that enhance the real-world usability of generative music systems. Our first design decision is to use the General MIDI format as input and output given that it is the most supported symbolic music encoding standard. We consider two main categories: I/O specifications, which place restrictions on the musical material that can be processed and generated by the system; and generation methods, favoring using existing musical content as prompt vs. unconditioned generation.

\subsection{Input/Output Specification}
We first define a track $(t_{\texttt{inst}})$, which is a distinct set of musical material (i.e. notes) played by an instrument ($\texttt{inst}$). In some cases, a track may be distinguished by its musical purpose (ex. $t_{\texttt{melody}}$) rather than its instrument. For example, MusicVAE~\cite{roberts2018hierarchical} aggregates all melodies into a single track type, rather than distinguishing between melodies based on their instrumentation (piano, synth, saxophone, etc.). In what follows, $t^m_{\texttt{inst}}$, $t^p_{\texttt{inst}}$ and $t_{\texttt{drum}}$ denote a monophonic track, a polyphonic track, and a drum track, respectively. For example, $t^m_{\texttt{bass}}$ denotes a monophonic bass track.

A generated excerpt can be described by a list of track types (ex. $[t^m_{\texttt{bass}},t^p_{\texttt{piano}}]$), and thus, we can define the output specification $(O^{\star})$ for an arbitrary generative music system as a set of track lists (ex. $O^{\star} = \{[t^m_{\texttt{bass}},t^p_{\texttt{piano}}],[t^p_{\texttt{piano}}, t^p_{\texttt{synth}}]\}$)
. We consider a system to have a fixed schema when $O^{\star}$ contains a single tracklist. For example, CoCoNet~\cite{roberts2018hierarchical} has a fixed schema, as $O^{\star} = \{[t^m_{\texttt{soprano}}, t^m_{\texttt{alto}}, t^m_{\texttt{tenor}}, t^m_{\texttt{bass}}]\}$, meaning that the system is only capable of generating 4 track music containing soprano, alto, tenor and bass tracks.

\begin{table*}[t]
    \centering
    \setlength{\tabcolsep}{1mm}
    \fontsize{9pt}{9pt}\selectfont
    \begin{tabular}{cccccccc}
    \toprule
    & \multicolumn{5}{c}{I/O Specifications} & \multicolumn{2}{c@{}}{Generation Tasks} \\
    \cmidrule(l){2-6} \cmidrule(l){7-8}

    System & \begin{tabular}{@{}c@{}}Number \\ of Tracks\end{tabular} & \begin{tabular}{@{}c@{}}Number of \\ Instruments\end{tabular} & \begin{tabular}{@{}c@{}}Fixed \\ Schema\end{tabular} & Drums & \begin{tabular}{@{}c@{}}Track-Level \\ Polyphony\end{tabular} & Infilling & \begin{tabular}{@{}c@{}}Attribute \\ Control\end{tabular} \\
    \midrule
    \rule{0pt}{4ex} {\bf \systemName} & {\bf Any} & {\bf 128} & {\bf no} & {\bf yes} & {\bf yes} & {\bf yes} & {\bf yes} \\
    FIGARO \cite{vonrutte2022figaro} & 128 & 128 & no & yes & yes & no & no \\
    MMT \cite{dong2023mmt} & 64 & 64 & yes & yes & yes & no & no \\
    MuseNet \cite{payne2019} & 10 & 10 & no & yes & yes & no & yes \\
    MuseGAN \cite{hong2019generative} & 4 & 4 & yes & yes & yes & no & no \\
    LahkNES \cite{donahue2019lakhnes} & 4 & 4 & yes & yes & no & no & no \\
    CoCoNet \cite{huang2019counterpoint} & 4 & 4 & yes & no & no & yes & no \\
    MusicVAE \cite{roberts2018hierarchical} & 3 & 3 & yes & yes & no & no & no \\
    MusIAC \cite{guo2022musiac} & 3 & 3 & yes & no & yes & yes & yes \\
    SketchNet \cite{DBLP:conf/ismir/0021WBD20} & 1 & no & yes & no & no & yes & yes \\
    \cite{pati2019learning},\cite{gautam_mittal_2021_5624363} & 1 & no & yes & no & no & yes & no \\
    \cite{DBLP:conf/ismir/ChangLY21},\cite{wayne_chi_2020_4245578} & 1 & no & yes & no & yes & yes & no \\
    \cite{DBLP:conf/ismir/TanH20},\cite{DBLP:conf/ismir/0008X21},\\\cite{DBLP:conf/ismir/0008WZX20} & 1 & no & yes & no & yes & no & yes\\
    \cite{haki_2022_7088343drumaccompaniment}, \cite{Nuttall2021TransformerBumbleBeat} & 1 & no & yes & yes & yes & no & no \\
    \cite{huang2018music} & 1 & no & yes & no & yes & no & no \\

    \bottomrule
    \end{tabular}
    \caption{A summary of the I/O specifications and generation tasks of recently published generative music systems.}
    \label{model_table}
\end{table*}

In Table \ref{model_table}, we describe the following features of symbolic music generation systems: the number of tracks, the number of instruments, whether a fixed schema of instruments is assumed, support for drum tracks, and support for polyphony at the track-level. Clearly, reducing the restrictions on a system's output increases the usability of a system, as it can accommodate a greater number of practices and user workflows. We leave aside style-specific generative systems \cite{ren2020popmag,collins_2023_10113434expressor,wu2020jazztransformer,liu2022symphonynet,huang2020pop,hsiao2021compound}, as we aim for a style-agnostic system that can accommodate as many users as possible.

As shown in Table \ref{model_table}, most systems either support a single track or require a fixed schema of instruments. One exception is MuseNet~\cite{payne2019}, which supports up to 10 tracks and any subset of the 10 available instruments. However, there are significant differences between MuseNet and \systemName. MuseNet uses separate \TT{NOTE\_ON} and \TT{NOTE\_OFF} tokens for each pitch on each track, placing inherent limitations on the number of tracks that can be represented, as the token vocabulary size cannot grow unbounded\cite{bpe-for-symbolique-music2023}. Considering that MuseNet is currently the largest model in terms of the number of weights, the number of tracks is unlikely to be increased without altering the representation. Instead, we decouple track information from \TT{NOTE\_ON}, \TT{NOTE\_DUR}, and \TT{NOTE\_POS} tokens, allowing the use of the same tokens for each track. Although this is a relatively small change, it enables us to accommodate all 128 General MIDI instruments. Furthermore, there is no inherent limit on the number of tracks, as long as the entire $n$-bar multi-track sequence can be encoded using less than 2048 tokens. Practically, this means more than 10 tracks can be generated at once depending on their content. Note that the upper limit of 2048 tokens is not a limitation of the representation itself, but rather the size of the model, and this limitation could be addressed with larger and more memory-intensive models. Both MuseNet and \systemName\, do not require a fixed instrument schema, however, MuseNet treats instrument selections as a suggestion, while \systemName\, guarantees a particular instrument will be used.

\subsection{Generation Tasks}

We consider four different generation tasks: unconditional generation, continuation, infilling, and attribute control. Unconditional generation produces music from scratch. Besides changing the data that the model is trained on, the user has limited control over the output of the model. Continuation involves conditioning the model with musical material temporally preceding the music that is to be generated. Since both unconditional generation and continuation come for free with any auto-regressive model trained on a temporally ordered sequence of musical events, all systems are capable of generating musical material in this manner. Infilling conditions generation on a subset of musical material, asking the model to fill in the blanks, so to speak. Although the terms infilling and inpainting are often used interchangeably, some important distinctions must be made in our context. In contrast to inpainting a section of an image, where the exact location and number of pixels to be inpainted are defined before generation, when infilling a section of music the number of tokens to be generated is unknown. Furthermore, in the context of multi-track music that is represented using a single time-ordered sequence where tracks are interleaved, the location of tokens to be added is unknown. This makes bar-level and track-level infilling quite complex, directly motivating the representation we describe in Section \ref{propRep}. With tracks ordered sequentially, the location of the tokens to be infilled is then known. Infilling can occur at different levels (i.e. note-level, bar-level, track-level). Track-level infilling is the most coarse and allows a set of $n$-tracks to be generated that are conditioned on a set of $k$ existing tracks, resulting in a composition with $k + n$ tracks. Bar-level infilling allows for $n$-bars selected across one or more tracks to be re-generated, and conditioned on the remaining content - past, current, and future - both on the track(s), and all other tracks.

\section{Proposed Music Tokenization}\label{propRep}

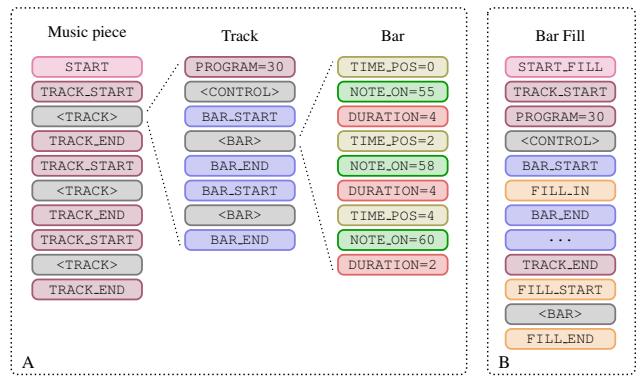
\begin{figure}[t!]
    \centering
    \resizebox{\columnwidth}{!}{
    \begin{tikzpicture}[
        token/.style={very thick,align=center,rounded corners=0.4em, minimum width=7.4em, minimum height=1.35em},
        stacked_tok/.style={token, below=0.2em of 1},
        red_tok/.style={draw=myred!60,fill=myred!20},
        blue_tok/.style={draw=myblue!50,fill=myblue!20},
        green_tok/.style={draw=mygreen,fill=mygreen!20},
        yellow_tok/.style={draw=myyellow,fill=myyellow!20},
        orange_tok/.style={draw=myorange!60,fill=myorange!20},
        purple_tok/.style={draw=mypurple!65,fill=mypurple!20},
        pink_tok/.style={draw=mypink!65,fill=mypink!20},
        gray_tok/.style={draw=mygray!65,fill=mygray!20},
        ]

        \node [token, pink_tok] (1) at (0,0) {\texttt{START}}; 
        \node (ref_multitrack) at (1) {};
        \node [stacked_tok, purple_tok] (1) {\texttt{TRACK\_START}};
        \node [stacked_tok, gray_tok] (1) {\texttt{<TRACK>}};
        \node [right=3.7em of 1] (ref_track_tok) at (1) {};
        \node [stacked_tok, purple_tok] (1) {\texttt{TRACK\_END}};
        \node [stacked_tok, purple_tok] (1) {\texttt{TRACK\_START}};
        \node [stacked_tok, gray_tok] (1) {\texttt{<TRACK>}};
        \node [stacked_tok, purple_tok] (1) {\texttt{TRACK\_END}};
        \node [stacked_tok, purple_tok] (1) {\texttt{TRACK\_START}};
        \node [stacked_tok, gray_tok] (1) {\texttt{<TRACK>}};
        \node [stacked_tok, purple_tok] (1) {\texttt{TRACK\_END}};
        \node [align=center, above=1.4em of ref_multitrack, anchor=south] (1) at (ref_multitrack) {\large Music piece};

        \node [token, purple_tok, right=6.2em of ref_multitrack] (1) {\texttt{PROGRAM=30}};
        \node (ref_track) at (1) {};
        \node [left=3.7em of 1] (ref_track_line1) at (1) {};
        \node [stacked_tok, gray_tok] (1) {\texttt{<CONTROL>}};
        \node [stacked_tok, blue_tok] (1) {\texttt{BAR\_START}};
        \node [stacked_tok, gray_tok] (1) {\texttt{<BAR>}};
        \node [right=3.7em of 1] (ref_bar_tok) at (1) {};
        \node [stacked_tok, blue_tok] (1) {\texttt{BAR\_END}};
        \node [stacked_tok, blue_tok] (1) {\texttt{BAR\_START}};
        \node [stacked_tok, gray_tok] (1) {\texttt{<BAR>}};
        \node [stacked_tok, blue_tok] (1) {\texttt{BAR\_END}};
        \node [left=3.7em of 1] (ref_track_line2) at (1) {};
        \node [align=center, above=1.4em of ref_multitrack, anchor=south] (1) at (ref_track) {\large Track};

        \node [token, yellow_tok, right=6.2em of ref_track] (1) {\texttt{TIME\_POS=0}};
        \node (ref_bar) at (1) {};
        \node [left=3.7em of 1] (ref_bar_line1) at (1) {};
        \node [stacked_tok, green_tok] (1) {\texttt{NOTE\_ON=55}};
        \node [stacked_tok, red_tok] (1) {\texttt{DURATION=4}};
        \node [stacked_tok, yellow_tok] (1) {\texttt{TIME\_POS=2}};
        \node [stacked_tok, green_tok] (1) {\texttt{NOTE\_ON=58}};
        \node [stacked_tok, red_tok] (1) {\texttt{DURATION=4}};
        \node [stacked_tok, yellow_tok] (1) {\texttt{TIME\_POS=4}};
        \node [stacked_tok, green_tok] (1) {\texttt{NOTE\_ON=60}};
        \node [stacked_tok, red_tok] (1) {\texttt{DURATION=2}};
        \node [left=3.7em of 1] (ref_bar_line2) at (1) {};
        \node [align=center, above=1.4em of ref_multitrack, anchor=south] (1) at (ref_bar) {\large Bar};

        \node [token, pink_tok, right=7.2em of ref_bar] (1) {\texttt{START\_FILL}};
        \node (ref_bar_fill) at (1) {};
        \node [stacked_tok, purple_tok] (1) {\texttt{TRACK\_START}};
        \node [stacked_tok, purple_tok] (1) {\texttt{PROGRAM=30}};
        \node [stacked_tok, gray_tok] (1) {\texttt{<CONTROL>}};
        \node [stacked_tok, blue_tok] (1) {\texttt{BAR\_START}};
        \node [stacked_tok, orange_tok] (1) {\texttt{FILL\_IN}};
        \node [stacked_tok, blue_tok] (1) {\texttt{BAR\_END}};
        \node [stacked_tok, blue_tok] (1) {\texttt{...}};
        \node [stacked_tok, purple_tok] (1) {\texttt{TRACK\_END}};
        \node [stacked_tok, orange_tok] (1) {\texttt{FILL\_START}};
        \node [stacked_tok, gray_tok] (1) {\texttt{<BAR>}};
        \node [stacked_tok, orange_tok] (1) {\texttt{FILL\_END}};

        \node [align=center, above=1.4em of ref_multitrack, anchor=south] (1) at (ref_bar_fill) {\large Bar Fill};

        \draw[dotted, line width=0.3mm](ref_track_tok.north)--(ref_track_line1.north);
        \draw[dotted, line width=0.3mm](ref_track_tok.south)--(ref_track_line2.south);

        \draw[dotted, line width=0.3mm](ref_bar_tok.north)--(ref_bar_line1.north);
        \draw[dotted, line width=0.3mm](ref_bar_tok.south)--(ref_bar_line2.south);

        \draw[dotted, line width=0.35mm, rounded corners] (-1.8, 1.4) rectangle (9, -7.3);
        \draw[dotted, line width=0.35mm, rounded corners] (9.5, 1.4) rectangle (13, -7.3);
        \node [align=center] (1) at (-1.4, -7.0) {\Large A};
        \node [align=center] (1) at (9.9, -7.0) {\Large B};

    \end{tikzpicture}
    }
    \caption{The Multitrack (A) and Bar-Fill (B) tokenizations. The grey \TT{<BAR>}, \TT{<TRACK>} and \TT{<CONTROL>} placeholders correspond token subsequences of complete bars, complete tracks, and attribute controls, respectively.}
    \label{fig:REP}
\end{figure}

In this section, we introduce two tokenizations to interpret musical compositions: the Multi-Track representation and the Bar-Fill representation. In contrast to other systems~\cite{oore2018time,huang2018music}, which use \TT{NOTE\_ON}, \TT{NOTE\_OFF} and \TT{TIME\_DELTA} tokens, we represent musical material using an approach which was previously employed for the Pop Music Transformer~\cite{huang2020pop}. In our Multi-Track representation, each bar of music is represented by a sequence of tokens, which include:
\begin{itemize}
\item 128 \TT{NOTE\_ON} tokens: These represent the pitch of each note in the bar.
\item 96 \TT{TIME\_POSITION} tokens: These represent the absolute start time (the time elapsed since the beginning of the bar, as opposed to the time elapsed since the last event) of each note within the bar.
\item 96 \TT{DURATION} tokens: These represent the duration of each note. Both the \TT{DURATION} and \TT{TIME\_POSITION} tokens range from a sixteenth-note triplet to a double whole note in sixteenth-note triplet increments.

\end{itemize}

We delimit a bar with \TT{BAR\_START} and \TT{BAR\_END} tokens. A sequence of bars makes up a track, which is delimited by \TT{TRACK\_START} and \TT{TRACK\_END} tokens. At the beginning of each track, one of 128 \TT{INSTRUMENT} token specifies its MIDI program. Tokens that condition the generation of each track on various musical attributes follow the \TT{INSTRUMENT} token, and will be discussed in Section \ref{attrctrl}. The tracks are then nested within a multi-track piece, which begins with a \TT{START} token. Note that all tracks are played simultaneously, not sequentially. This process of nesting bars within a track and tracks within a piece is illustrated in Figure \ref{fig:REP}A. Notably, we do not use an \TT{END} token, as we can simply sample until we reach the $n^{th}$ \TT{TRACK\_END} token if we wish to generate n tracks. This tokenization is implemented in MidiTok \cite{miditok2021} for ease of use.

The Multi-Track representation allows the model to condition the generation of each track on the tracks that precede it, which allows for a subset of the musical material to be fixed while generating additional tracks. However, this representation doesn't provide control at the bar level, except in cases where the model is asked to complete the remaining bars of a track. In other words, the model cannot fill in bars that are in the middle of a track. To generate a specific bar in a track conditioned on the other bars, we introduce the Bar-Fill representation. In this representation, bars to be predicted are replaced by a \TT{FILL\_IN} token. These bars are then placed/generated at the end of the piece after the last track token, and each bar is delimited by \TT{FILL\_START} and \TT{FILL\_END} tokens (instead of \TT{BAR\_START} and \TT{BAR\_END} tokens). 

Note that during training, the bars with \TT{FILL\_IN} tokens appear in the same order as they appeared in the original Multi-Track representation, shown in Figure \ref{fig:REP}B. By ordering the bars consistently, the model learns to always output tokens in the same order as the bars that are marked for generation. The Bar-Fill representation begins with a \TT{START\_FILL} token instead of a \TT{START} token. The Multi-Track representation is simply a special case of the Bar-Fill representation, where no bars are selected for infilling.

\subsection{Adding Interpretation Expressiveness}

Multiple attempts at generating expressive symbolic music have been made either as an independent process \cite{pmlr-v97-gillick19a, cancino2016nlbm, malik2017neural, Maezawa2019RenderingMP} or a simultaneous process with the musical content generation \cite{oore2018time, huang2018music, hawthorne2018nade, huang2020pop, yusong2021mididdsp}. None, however, allow for expressive multitrack generation. Here, we focus on velocy, as a proxy for dynamics, and microtiming, as the two main aspects of expressive music interpretation. We implement two extensions to our current tokenisation allowing the simultaneous generation of expressive MIDI. This allows us to leverage the 31\% of MIDI files in GigaMIDI that have been marked as expressive (varying velocity, and non-quantized micro-timing).

Firstly, we include 128 \TT{VELOCITY} tokens that encode every possible velocity level of a MIDI note, as velocity is a proxi for dynamics and an important aspect of expressiveness in musical performances.

\begin{figure}[!t]
\centering
\begin{subfigure}[t]{.45\columnwidth}
    \centering
        \resizebox{\columnwidth}{!}{
         \begin{tikzpicture}
        \node (start_graph) at (0,0) {};
        \node (end_graph) at (5.3,0) {};
        \draw [>=latex,->] (0,0) -- (end_graph.west);
        \draw (0,-0.2) -- (0,0) ;
        \draw [dashed] (0,0) -- (0,1.8) ;
        \draw (2,-0.2) -- (2,0) ;
        \draw [dashed] (2,0) -- (2,1.8) ;
        \draw (4,-0.2) -- (4,0) ;
        \draw [dashed] (4,0) -- (4,1.8) ;
        \draw [dotted] (0.5,1.6) -- (0.5,-0.2) ;
        \draw [draw=mygray!65,fill=mygray!20,rounded corners=0.4em,very thick]  (0.5, 1.6) rectangle (5.3, 0.8) ;
        \node at (2.9,1.16) [font=\Large] {Pitch=55, Velocity=98};
        \node at (0.8, 0.37) [font=\LARGE] {$t_s$};
        \node at (2.3, 0.37) [font=\LARGE] {$t_k$};
        \draw [>=latex,<->] (0.5, -0.22) -- (1.98, -0.22);
        \node at (1.6, -0.68) [font=\LARGE] {$\Delta t = t_s - t_k$};

        \node at (0, 2.1) [font=\LARGE] {k-1};
        \node at (2, 2.1) [font=\LARGE] {k};
        \node at (4, 2.1) [font=\LARGE] {k+1};
        
        \end{tikzpicture}
        }
         \caption{Time difference between the note onset and the quantized \TT{TIME\_POS} value.}
         \label{fig:midi_m}
\end{subfigure}
\hspace{0.01\textwidth}
\begin{subfigure}[t]{.45\columnwidth}
   \centering
        \resizebox{\columnwidth}{!}{
        \begin{tikzpicture}[
            token/.style={very thick,align=center,rounded corners=0.4em, minimum width=8.7em, minimum height=1.35em, font=\Large},
            stacked_tok/.style={token, below=0.2em of 1},
            red_tok/.style={draw=myred!60,fill=myred!20},
            blue_tok/.style={draw=myblue!50,fill=myblue!20},
            green_tok/.style={draw=mygreen,fill=mygreen!20},
            yellow_tok/.style={draw=myyellow,fill=myyellow!20},
            orange_tok/.style={draw=myorange!60,fill=myorange!20},
            purple_tok/.style={draw=mypurple!65,fill=mypurple!20},
            pink_tok/.style={draw=mypink!65,fill=mypink!20},
            gray_tok/.style={draw=mygray!65,fill=mygray!20},
            ]
    
            \node [stacked_tok, yellow_tok] (1) at (0,0) {\texttt{TIME\_POS=k}}; 
            \node (ref_score_toks) at (1) {};
            \node [right=4em of 1] (ref_start) at (1) {};
            \node [stacked_tok, gray_tok] (1) {\texttt{<DELTA>}};
            \node [right=4em of 1] (ref_delta_toks) at (1) {};
            \node [stacked_tok, green_tok] (1) {\texttt{NOTE\_ON=55}};
            \node [stacked_tok, orange_tok] (1) {\texttt{VELOCITY=98}};
            \node [stacked_tok, red_tok] (1) {\texttt{DURATION=1}};
            \node [align=center, above=1.4em of ref_score_toks, anchor=south] (1) at (ref_score_toks) [font=\LARGE] {Note Tokens};
    
            \node [token, purple_tok, right=1.8em of ref_start] (1) {\texttt{DELTA=-1}};
            \node (delta_tok) at (1) {};
            \node [left=4em of 1] (ref_track_line1) at (1) {};
            \node [stacked_tok, purple_tok] (1) {\texttt{DELTA=$|\Delta t|$}};
            \node [left=4em of 1] (ref_track_line2) at (1) {};
            \node [align=center, above=1.4em of ref_score_toks, anchor=south] (1) at (delta_tok) [font=\LARGE] {Delta tokens};

            \draw[dotted, line width=0.3mm](ref_delta_toks.north)--(ref_track_line1.north);
            \draw[dotted, line width=0.3mm](ref_delta_toks.south)--(ref_track_line2.south);
    
    
         \end{tikzpicture}
         }
         \caption{Token subsequence corresponding to a note, with microtiming and velocity.}
         \label{fig:with_m}
\end{subfigure}
\caption{Adding expressivity to the token sequence}
        \label{fig:tokens_micro}
\end{figure}
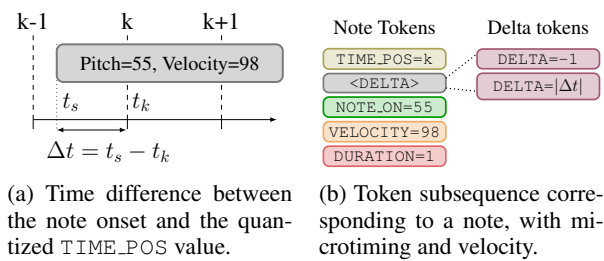

Secondly, we include new tokens to represent microtiming. Our current tokenization allows for 96 different \TT{TIME\_POSITION} tokens within a bar. Therefore, this level of quantization occurs in the model which does not capture microtiming. Intuitively, a solution to this problem would be to increase the vocabulary and time resolution of the \TT{TIME\_POSITION} tokens. However, to maintain the possibility of using the current downsampled and non-expressive tokenization while allowing the possibility to add expressiveness, we introduce a new token \TT{DELTA} which encodes the time difference of the original MIDI note onset $t_s$ from the quantized token onset $t_k$, illustrated in Figure \ref{fig:midi_m}. The \TT{DELTA} tokens encode the offset in increments of 1/160th of a sixteenth-note triplet. We consider  80 additional tokens because the maximal absolute time difference is half of a sixteenth-note triplet, and we use a \TT{DELTA\_{-1}} token when this time difference is negative. This resolution allows for a small addition to the vocabulary, yet is enough to encode 99\% of the expressive tracks of GigaMIDI.
The use of expressive tokens is illustrated in Figure \ref{fig:with_m}.

\section{Controlling Music Generation} \label{attrctrl}

The premise behind attribute controls is that given a musical excerpt $x$, and a measurable musical attribute $a$ for which we can compute a categorical or ordinal value from $x$ (i.e. $a(x)$), the model will learn the conditional relationship between tokens representing $a(x)$ and the musical material on a track, provided these tokens precede the musical material. Practically, this is accomplished by inserting one or more \TT{CONTROL} tokens which specify the level of a particular musical attribute $a(x)$ immediately after the \TT{INSTRUMENT} token (see Figure \ref{fig:REP}), and before the tokens which specify the musical material. As a result, our approach is most certainly not limited to the specific musical attributes we discuss below, and can be applied to control any musical feature that can be measured. We employ three approaches to control musical attributes of the generated material: categorical controls, which condition generation on one of $n$ different categories; value controls, which condition generation on one of $n$ different ordinal values; and range controls, which condition the system to generate music wherein a particular musical attribute has values that fall within a specified range. 

Instrument control is an example of a categorical control, as one of 128 different instrument types can be selected. We use a value control for note density, however, the density categories are determined relative to the instrument type, as average note density varies significantly between instruments. For each of the 128 general MIDI instruments, we calculate the number of note onsets for each bar in the dataset. We divide the distribution for each instrument $\sigma$ into 10 regions with the range $[\PP{10i},\PP{10(i+1)})$ for $0 \leq i < 10$, where $\PP{n}$ denotes the $n^{th}$ percentile of the distribution $\sigma$. Each region corresponds to a different note density level for a particular instrument.

We choose to apply range controls to note duration and polyphony. Each note duration $(d)$ is quantized as $\lfloor\log_2(d)\rfloor$. The quantization process groups note durations into 5 different bins $[\frac{1}{32},\frac{1}{16}), [\frac{1}{16},\frac{1}{8}), [\frac{1}{8},\frac{1}{4}), [\frac{1}{4},\frac{1}{2})$, and $[\frac{1}{2},\frac{1}{1})$, which we will refer to as note duration levels. Then the $15^{th}$ and $85^{th}$ percentiles of a distribution containing all note duration levels within a track are used to condition generation. Polyphony levels follow a similar approach. The number of notes simultaneously sounding (i.e. polyphony level) at each timestep is calculated (a timestep is one $16^{th}$ note triplet). Then we use the $15^{th}$ and $85^{th}$ percentiles of a distribution containing all polyphony levels within a track. For both these controls, we use two tokens, one to specify the lower bound and another for the upper bound. Admittedly, this is fuzzy range control, as strict range control would typically use the smallest and largest values in the distribution ($0^{th}$ and $100^{th}$ percentiles respectively). We elected to use the $15^{th}$ and $85^{th}$ percentiles in order to mitigate the effect of outliers within the distribution, decreasing the probability of exposing the model to ranges in which values are heavily skewed to one side of the range.

\section{Training \systemName}

We use the new GigaMIDI \cite{Gigamidi2024} dataset, which builds on the MetaMIDI dataset \cite{jeffrey_ens_2021_metamidi}, to train with a split of: $p_{train}=80\%$, $p_{valid}=10\%$, and $p_{test}=10\%$. Our model is built on the GPT2 architecture \cite{radford2019language}, implemented using the HuggingFace Transformers library \cite{Wolf2019HuggingFacesTS}. The configuration of this model includes 8 attention heads and 6 layers, utilizing an embedding size of 512 and an attention window encompassing 2048 tokens. This results in approximately 20 million parameters. 

For each batch, we pick 32 random MIDI files (batch size) from the respective split of the dataset (train, test, valid) and pick random 4 or 8-bar multi-track segments from each MIDI file. For a segment with $n$ tracks, we pick $k$ tracks randomly selecting a value for $k$ on the range $[2,\min(n,12)]$. With 75\% probability, we do bar infilling on a segment where we mask up to 75\% of the bars. The number of bars is selected uniformly from values on the range $[0, \lfloor{n_{tracks} * n_{bars} * 0.75}\rfloor]$. Then, we randomly transpose the musical pitches (except for the drum track, of course) with a value for the range $[-6,5]$. Each time we select a $n$-bar segment during training, we randomly order the tracks so that the model learns each possible conditional ordering between different types of tracks. The model is trained to predict bar, track, and instrument tokens. As a result, when generating a new track, the model can select a sensible instrument to accompany the pre-existing tracks, thus learning instrumentation.

We train with the Adam optimizer, a learning rate of $10^{-4}$, without dropout. Training to convergence typically takes 2-3 days using 4 V100 GPUs. We pick the model with the best validation loss.

\begin{figure}[t!]
    \centering
    \begin{subfigure}{8.2cm}
    \centering
        \includegraphics[width=\columnwidth]{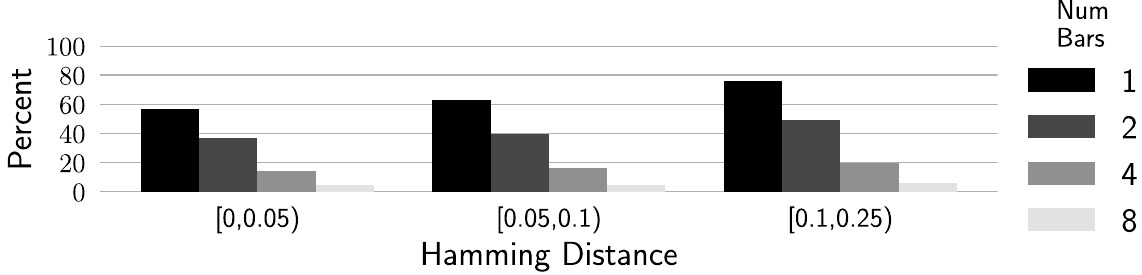}
        \caption{Hamming distance}
        \label{fig:plag_corpus}
    \end{subfigure}
    \begin{subfigure}{8.2cm}
        \centering
        \includegraphics[width=\columnwidth]{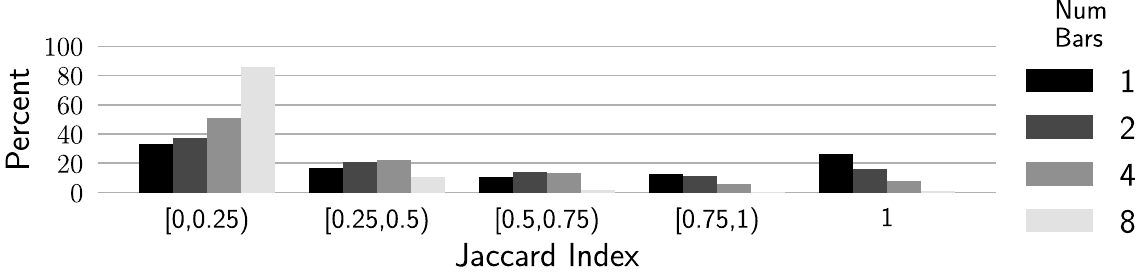}
        \caption{Jaccard Index}
        \label{fig:plag_1}
    \end{subfigure}
    \caption{The percentage of generated excerpts $(g_i)$ with a Hamming distance (resp. Jaccard Index $\mathcal{J}(o_i,g_i)$) between any excerpt $o_i$ from the training dataset and $g_i$ on the range [a,b). A Hamming distance (resp. Jaccard Index) of distance 0 (resp. of 1) indicates two excerpts are identical, while 1 (resp. 0) indicates they are very different.}
\end{figure}

\section{Sampling with \systemName}
To achieve syntactically valid outputs from the system (with respect to the tokenization used), we incorporate specific masking constraints. More precisely, we mask select tokens during various stages of the model's inference process to preserve the sequence necessary for encoding, decoding, and prediction tasks. For example, a scenario in which a \TT{BAR\_END} token appears after a \TT{NOTE\_ON} token token is not feasible, given that the \TT{DURATION} and \TT{NOTE\_POSITION} have yet to be determined. In such instances, we mask the \TT{BAR\_END} to prevent its sampling. Subsequently, we sample among the remaining unmasked tokens. This rule-based sampling approach ensures that the model maintains a logical structure throughout its operations.

\section{ Release, Evaluation, and Applications }
\systemName\, has been released\footnote{\url{https://www.metacreation.net/projects/mmm} links to models and various examples of generations.} and is seeing real-world usage in several contexts, which directly supports our assertion that \systemName\, is a practical model for computer-assisted composition. There are ongoing collaborations for the integration of \systemName\, into synthesizers, like the OP-Z by Teenage Engineering\footnote{\url{https://www.metacreation.net/projects/opz-mmm}}, and game music composition software by Elias. \systemName\, has been integrated into the Cubase\footnote{\url{www.metacreation.net/projects/mmm-cubase}} digital audio workstation, the Calliope\footnote{\url{https://www.metacreation.net/projects/calliope}} web application, and an Ableton\footnote{\url{https://www.metacreation.net/projects/mmm4live}} plugin has been developed. \systemName\, has been used to compose music\footnote{\url{https://www.metacreation.net/projects/mmm-music}}, including two entries to the 2022 and 2023 AI Song contests, four albums (two by Philip Tremble, one by Monobor, and one by a selection of American and Canadian composers). It has been used to compose adaptive music for games\cite{videogameMMM}, and a yearly artistic residency\footnote{\url{https://vancouver.consulfrance.org/Artificial-Muse-residency-Vancouver}} with French composers is ongoing.

A user study \cite{renaudMMM} was conducted to evaluate the integration of a previous version of \systemName\, into a popular digital audio workstation. The study measured usability, user experience, and technology acceptance for two groups of experienced composers: hobbyists and professionals with convincing results. Since we have already conducted a comprehensive user study, we do not repeat a listening study here. Instead, our experiments are designed to address other aspects that impact the usability of the system in real-world settings.

In the following, our evaluation of \systemName\, gauges the performance of the system by addressing the following research questions:
\begin{enumerate}
    \item Originality: Does \systemName\, generate original variations or simply duplicate material from the dataset?
    \item Stylistic Similarity: Does \systemName\, generate musical material that is stylistically similar to the dataset (i.e., well-formed music)?
    \item Attribute Controls: How effective are density level, polyphony range, and note duration range controls?
\end{enumerate}

\subsection{Evaluating the Originality of Generated \protect\\
Material}\label{origexp}

\subsubsection{Intra-Dataset Originality} \label{corp_orig}

It is increasingly important to quantify the frequency with which a generative system is producing musical material that is nearly identical to the training dataset, given potential legal issues that may arise when these systems are deployed into the real world, and the difficulty of guaranteeing that a generative system does not engage in this type of behavior \cite{DBLP:conf/aaai/PapadopoulosRP14}. To accomplish this, we represent musical material in each track as a piano roll and use hamming distance to calculate the distance between two piano rolls. A piano roll is a $T \times 128$ boolean matrix specifying when particular pitches are sounding, where $T$ is the number of time-steps. Note that when we calculate Hamming distance between two piano rolls, we normalize the distance by the size of the piano roll. Therefore, the distance between maximally different piano rolls would be 1.

Since the dataset contains hundreds of thousands of unique MIDI files, we are faced with a time complexity issue, and must employ some heuristics to speed this process up. First, rather than searching nearly identical $n$-bar piano rolls, we search using single-bar piano rolls, and aggregate the results of $n$ search processes. Note that this means that if $n-1$ of the bars have a match in the dataset, but one of the bars does not, the $n$-bar excerpt will not be considered to have a match in the dataset. However, since we are interested in identifying nearly identical matches, this is unlikely to cause much of an issue. To filter out highly dissimilar candidate matches efficiently, we compute the Hamming distance between compressed piano rolls first. Given a $48 \times 128$ piano roll $x$ that represents a single 4/4 bar of musical material, we discard notes outside the range $[21,109)$ and take the maximum value over each consecutive set of 6 time-steps (equivalent to one 1/8 note) on the first axis, producing a $8 \times 88$ matrix $x$. We calculate the Hamming distance between the compressed piano rolls, discarding any candidate matches that have a distance greater than 0.25, and then compute Hamming distance on the full-sized piano rolls for the remaining candidate matches. Even with these optimizations, the search is executed in parallel on a 32-core machine and takes an average of 83 seconds to complete a search for a single 4-bar excerpt. In the worst case, it can take up to an hour for a single query. Although we would have preferred to use Jaccard index rather than Hamming distance, as we do in the next sub-section, the nature of the heuristics employed to increase computation speed prohibited this option.

We compute 100 trials where we randomly select a 4-track 8-bar musical segment from the test split of the dataset, blank out $n$ consecutive bars on a single track, and generate (i.e. infill) a new set of bars $(g_i)$. Given a Hamming distance threshold, we determine if $g_i$ is nearly identical to any $n$-bar excerpt in the training split of the dataset, using the method described above. In Figure \ref{fig:plag_corpus} we present the percentage of trials for which the Hamming distance between any excerpts in the training dataset and $g_i$ is on the specified range. Unsurprisingly, as the number of bars increases, the percentage of instances where \systemName\, duplicates the training data decreases significantly. This correlation was expected as shorter generations are more constrained by the surrounding musical content. There are not that many one-track, one-bar musical excerpts that are not already in the data set.

\subsubsection{Infilling Originality}
\label{infillorig}

We ought to also measure the data reproduction for the infilling task, which may occur when the model predicts the exact segment that a user wants to infill, resulting in no change and inevitable frustration from the user. To measure the frequency with which this occurs, we randomly select a 4 track 8 bar musical segment from the test split of the dataset, blank out $n$ consecutive bars on a single track $(o_i)$, and generate a new set of $n$ bars ($g_i$) to replace ($o_i)$. Then, we measure the Jaccard index between piano roll representations of $o_i$ and $g_i$. We repeat this process 250 times for each number of bars $(n = 1,2,4,8)$ and report the results in Figure \ref{fig:plag_1}. On the whole, as the number of bars increases, the frequency with which the original material is duplicated decreases. Taken collectively, the results in this section indicate that \systemName\, can reliably produce original variations when generating 4 or more bars. In practice, when deployed in products, it always does as we actually test that the material generated is different from what is being replaced, and regenerate otherwise. We also re-generate tracks or bar infilling resulting in silence, which the user never intends.

\subsection{Quantifying Stylistic Similarity}

\begin{figure}
    \centering
    \includegraphics[width=\columnwidth]{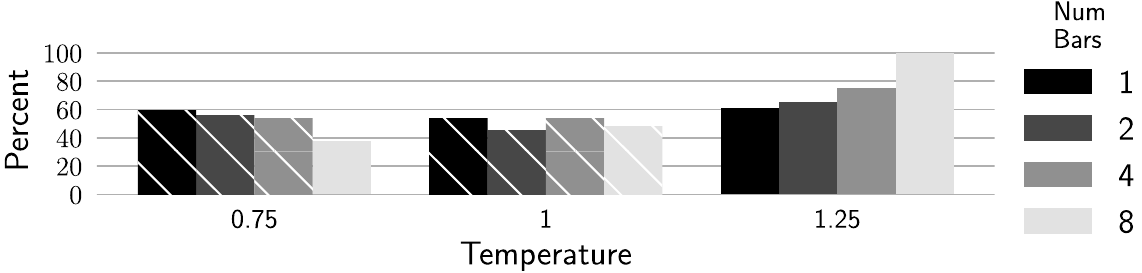}
    \caption{The percentage of trials where $\mathcal{S}_{\hat{\mathcal{O}}^{\star}_{25},\hat{\mathcal{G}}^{\star}_{25}}^{\mathcal{C}^{\star}_{50}}(\hat{\mathcal{O}}^{\star}_{25},\mathcal{C}^{\star}_{50}) \leq \mathcal{S}_{\hat{\mathcal{O}}^{\star}_{25},\hat{\mathcal{G}}^{\star}_{25}}^{\mathcal{C}^{\star}_{50}}(\hat{\mathcal{G}}^{\star}_{25},\mathcal{C}^{\star}_{50})$. Hatching indicates that the binomial test was insignificant, indicating that $\hat{\mathcal{O}}^{\star}$ is not more similar to $\mathcal{C}$ than $\hat{\mathcal{G}}^{\star}$.}
    \label{fig:sim_1}
\end{figure}

\begin{figure}
    \centering
    \includegraphics[width=\columnwidth]{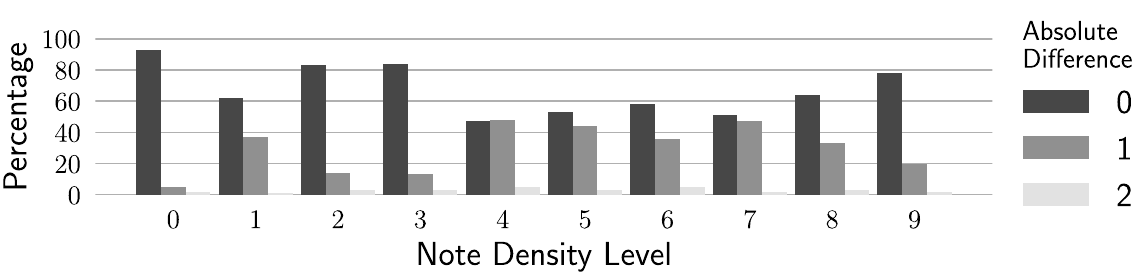}
    \caption{The percentage for each absolute difference between requested and actual note density.}
    \label{fig:attr_density}
\end{figure}

It is also important that the variations generated by the system are stylistically similar to the dataset. To be clear, we define musical style as the stylistic characteristics delineated by a set of musical data. As a result, when we claim to measure stylistic similarity to the training data, we are measuring similarity to the style that is delineated by this set of data. Consequently, we avoid having to make subjective decisions about what constitutes a particular musical style, while maintaining an evaluation framework that is generic enough to handle any arbitrary set of data.

We use \stylerank\,~\cite{ens2019eval} to measure the stylistic similarity of generated material. \stylerank\, is designed to measure the similarity of two or more groups of musical excerpts $(\mathcal{G}_1,...,\mathcal{G}_k)$ relative to a style delineated by a collection of ground truth musical excerpts $(\mathcal{C})$. Each musical excerpt is represented using a set of features, described in detail in the original paper, and a Random Forest classifier is trained to discriminate between $\mathcal{G}_1,...,\mathcal{G}_k$ and $\mathcal{C}$. Using an embedding space constructed from the trained Random Forest classifier, the average similarity between $\mathcal{G}_i$ and $\mathcal{C}$ can be computed for each $i$. In what follows, let $\mathcal{S}_{\mathcal{G}_1,...,\mathcal{G}_k}^{\mathcal{C}}(a,b)$ denote the median similarity between $a$ and $b$, calculated using a \stylerank\, instance trained on $\mathcal{G}_1,...,\mathcal{G}_k$ and $\mathcal{C}$.

For this experiment, we use the same musical excerpts from Section \ref{origexp} $(\mathcal{O} = \{o_1,...,o_{250}\}, \mathcal{G} = \{g_1,...,g_{250}\})$, however, we remove each pair $(o_i,g_i)$ where $\mathcal{J}(o_i,g_i) \geq 0.75$, producing $\hat{\mathcal{O}}$ and $\hat{\mathcal{G}}$. This ensures that we do not bias our measurements by including generated material that is nearly identical to the original preexisting material $(o_i)$ from the dataset. We also assemble a set of 1000 $n$-bar segments ($\mathcal{C}$) from the dataset. For each trial, we compute $\mathcal{S}_{\hat{\mathcal{O}}^{\star}_{25},\hat{\mathcal{G}}^{\star}_{25}}^{\mathcal{C}^{\star}_{50}}(\hat{\mathcal{O}}^{\star}_{25},\mathcal{C}^{\star}_{50}) \leq \mathcal{S}_{\hat{\mathcal{O}}^{\star}_{25},\hat{\mathcal{G}}^{\star}_{25}}^{\mathcal{C}^{\star}_{50}}(\hat{\mathcal{G}}^{\star}_{25},\mathcal{C}^{\star}_{50})$, where $X^{\star}_{n}$ denotes a subset of $X$ containing $n$ elements, which are selected randomly for each trial. In other words, for each trial, we determine if the median similarity between two subsets of the corpus is less than or equal to the median similarity between a subset of the corpus and a set of generated excerpts. We collect the results for 100 trials and compute a binomial test. If there is no significant difference between the count of trials for which the condition is true and the count of trials for which the condition is false, we can conclude that there is not a significant difference between the generated material and the corpus with respect to the similarity metric we are using. We report the results of this test using different numbers of bars and temperatures in Figure~\ref{fig:sim_1}.

\begin{figure}
    \centering
    \begin{subfigure}[b]{8.2cm}
        \includegraphics[width=\columnwidth]{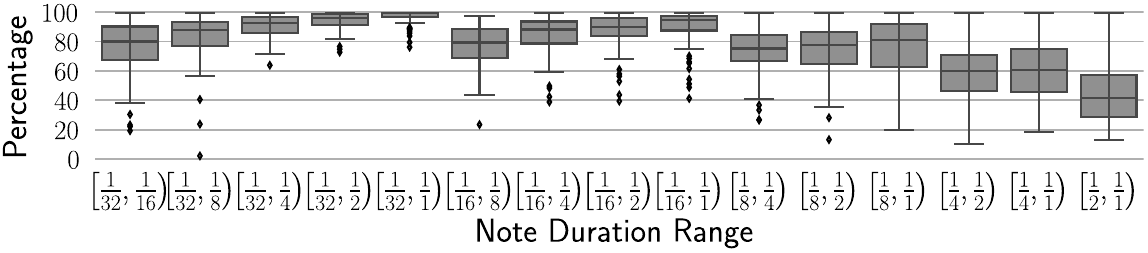}
        \caption{Note durations}
        \label{fig:attr_dur}
    \end{subfigure}
    \begin{subfigure}[b]{8.2cm}
        \centering
        \includegraphics[width=\columnwidth]{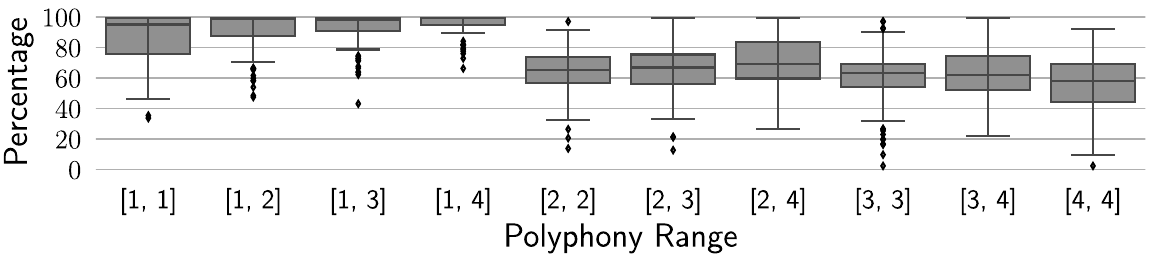}
        \caption{Polyphony levels}
    \label{fig:attr_poly}
    \end{subfigure}
    \caption{The percentage of note durations (a) and polyphony levels (b) within the range shown for 100 trials.}
\end{figure}

The results indicate that when generating with a temperature of 1.0, infilled generations are equivalent to the original preexisting material in terms of musical style (as quantified by \stylerank). Our results also show that when the temperature is greater than 1.0, the generated material is more frequently considered less similar to the dataset $(\mathcal{C})$ than $\hat{\mathcal{O}}$, an effect that increases along with the number of generated bars. This demonstrates that our measurement instrument is sensitive enough to detect small increases in the entropy of the music generated by the model, which are the byproduct of slightly changing the temperature.

This result does not only serve as an analytical evaluation of the model (as human evaluations are covered elsewhere \cite{renaudMMM}), but future work is to replicate these when conditioning on a set of files representing a given musical style. \systemName\ includes a categorical style control based on the style metadata extracted from the MetaMIDI dataset (a subset of GigaMIDI) and aligned to the MusicMap \cite{musicmap} ontology (Disco, Rock, Jazz, ...).

\subsection{Evaluating the Effectiveness of Attribute\protect\\Controls}

\systemName\, allows the user to condition generation not only on the existing musical content at the time of generation, but also on various control attributes such as the instrument, musical style, note density, polyphony level, and note duration. To evaluate how effective these control mechanisms are, we focus on the three last ones for brevity. We conduct 100 trials where we generate 8-bar segments from scratch using a particular attribute control method, and measure the difference between the anticipated outcome and the actual outcome. For note density control, we measure the absolute difference between the density level the generation was conditioned on and the density level of the generated material. For polyphony level and note duration, we compute the distribution of values (either polyphony level or note duration level) from the generated material, and count the percentage of values that fall within the specified range. If attribute control is successful, we would expect at least $70\%$ of the values to be within this range, as we used the $15^{th}$ and $85^{th}$ percentiles while training.

The results for note density, shown in Figure \ref{fig:attr_density}, demonstrate that the majority of times the absolute difference between the anticipated and actual density level is most often 0, and rarely exceeds 1. This indicates that this control method is effective. The results for Note Duration, shown in Figure \ref{fig:attr_dur} demonstrate that this attribute control method is quite effective, as the median outcome (in terms of percentage of note durations within the specified range) is at or above 70\% in all cases except for $[\frac{1}{4},\frac{1}{2})$, $[\frac{1}{4},\frac{1}{1})$ and $[\frac{1}{2},\frac{1}{1})$. In contrast, the Polyphony Level control is less effective, with the median outcome lying below the $70\%$ threshold in many cases. Calculating polyphony level at a single time-step is inherently more difficult than note duration, as the former requires knowledge of where multiple notes start and end, while the latter only requires knowing where one note starts and ends. This difference in difficulty seems to be reflected in the results, as \systemName\, is better at controlling note duration than polyphony.

In addition to these ``soft controls", we can also implement hard controls through rule-based sampling. This involves some bookkeeping. For example, for practical reasons, we also provide a hard polyphony limit $l_{poly}$. In places where it would otherwise be valid for a note to be inserted in the token sequence, we first check that the size of the set of currently sounding pitches $n_{pitch}$ satisfies $n_{pitch} < l_{poly}$. If $n_{pitch} = l_{poly}$, we mask note tokens.

\section{Conclusion}

We present \systemName, a style-agnostic generative system released as an Open RAIL-M licenced MMM model \cite{pasquier2018mmm}.  \systemName\, builds on an alternative approach to representing musical material, resulting in increased control over the generated output. We provided experimental evidence demonstrating the effectiveness of the system and outlined several ongoing real-world applications. The system runs on most personal computer with an attention window of 2048 tokens, coresponding to 8-16 bars depending on the number of tracks and their density. However, using an auto-regressive approach, and sliding windows, longer parts can be generated by repeatedly conditioning on portions of preexisting along with newly generated material. Future work involves: optimizing the model for real-time generation in musical agents, training larger models to expand the model attention window and attend to larger musical structures, expanding the set of attribute controls, and continuing integration of \systemName\, into real-world products and practices.

\section*{Ethical Statement}

While \systemName\, is style-agnostic, and does include more musical styles and content than any human can know, it still inherits from the dataset's bias, which only encompasses the musical styles afforded by the MIDI notation and available online. Arguably, there are still many musical styles under-represented (mostly non-western styles) or even absent from the dataset (i.e. not representable or available in MIDI) and therefore the model (although some suspecting users have noticed its generalization capabilities). Conversely, some musical styles are over-represented (e.g., pop). Further work is needed to qualify and quantify these biases.

Regarding copyright and intellectual property, MetaMIDI and GigaMIDI were acquired under the Fair Dealing law of Canada, which limits thei use to research and non-commercial use. We abide by these limitations.

Regarding \systemName\,: at the time of writing, it is unclear what the legal status of such a model is, as it does not contain the data itself, and produces original content for any non-over-constrained request (as shown above). So far, we restricted its use to non-commercial use. It is either released as-free-for use (Calliope), used for research purposes with collaborating companies, or used for research-creation purposes with selected artists. The ethical or legal implications of current creative use by other artists, and the potential use of the released model (under an Open RAIL-M licence) for commercial purposes by other parties does not rest with the authors of this research. We thus decline any responsibility for misuse.

\section*{Acknowledgments}
We would like acknowledge the support of the National Science and Engineering Research Council (NSERC), the Social Sciences and Humanities Research Council (SSHRC), the Canada Council for the Arts (CCA), and Steinberg Media Technology. We also thank Griffin Page and the anonymous reviewers for their edits and feedback.

\bibliography{aaai25}

\end{document}